\documentclass[12pt,pdf]{elsart} %geometry,
\usepackage{graphicx,amssymb,amsfonts,amsmath,bm,url} % see geometry.pdf on how to lay out the page. There's lots.

\newtheorem{remark}{Remark}
\newtheorem{theorem}{Theorem}
\def\pmatrix{\left(\begin{array}}
\def\endpmatrix{\end{array}\right)}
\def\RR{{\mathbb{R}}}
\def\bfc{{\bm{c}}}
\def\bfe{{\bm{e}}}
\def\bff{{\bm{f}}}
\def\bfv{{\bm{v}}}
\def\diag{{\rm diag}}

\begin{document}

\begin{frontmatter}

\title{Recent advances in bibliometric indexes \\ and the {\em PaperRank} problem\thanksref{t1}}
\thanks[t1]{Work developed within the project {\em ``Innovative Methods of Numerical Linear Algebra with Applications''}.}

\author[label1]{Pierluigi Amodio} \and
\ead{amodio@dm.uniba.it}
\author[label2]{Luigi Brugnano}
\ead{luigi.brugnano@unifi.it}

%\corauth[cor1]{Corresponding author.}
\address[label1]{Dipartimento di Matematica, Universit\`a di Bari, Italy}
\address[label2]{Dipartimento di Matematica ``U.~Dini'', Universit\`a di Firenze, Italy}

%\date{June 4, 2012} % delete this line to display the current date

\begin{abstract}
Bibliometric indexes are customary used in evaluating the impact of scientific research, even though it is very well known that in different research areas they  may range in very different intervals.  Sometimes, this is evident even within a single given field of investigation making very difficult (and inaccurate) the assessment of scientific papers. On the other hand, the problem can be recast in the same framework which has allowed to efficiently cope with the ordering of web-pages, i.e., to formulate the {\em PageRank} of Google. For this reason, we call such problem the {\em PaperRank} problem, here solved by using  a similar approach to that employed by {\em PageRank}. The obtained solution, which is mathematically grounded, will be used to compare the usual heuristics of the number of citations with a new one here proposed. Some numerical tests show that the new heuristics is much more reliable than the currently used ones, based on the bare number of citations. Moreo
 ver, we show that our model improves on recently proposed ones \cite{BiDCRo10}.
We also show that, once the {\em PaperRank} problem is correctly solved, one obtains, as a by-product, the ranking of both authors and journals. 
\end{abstract}


\end{frontmatter}

\section{Introduction}\label{uno}
In recent years, it has become the fashion to evaluate the impact of research by using bibliometric indexes. This approach clearly does not solve the problem at hand, though it can be useful to have a gross idea about specific issues. For instance, the quality of a scientist is sometimes ranked by using the so called {\em H-index} \cite{H05}, though it is very well known that this can be useful only for analysing short time return researches, whereas it could be completely inadequate to assess more basic research fields which, in turn, prove to be important (and, sometimes, priceless) only after decades or centuries. As an example, the {\em little Fermat theorem}, on which modern electronic secure transactions essentially rely, dates back to 1640, when it was apparently useless. Also, very important mathematicians are known to have very small {\em H-index} (e.g., Galois has an {\em H-index} equal to 2 \dots).

Nevertheless, in some circumstances, bibliometric indexes allow to obtain a rough idea about the impact of research, though their value heavily depends on the chosen index. In particular, it is recognized that the bare number of citations is a parameter which has many drawbacks: it does depend on the specific field of research, on unfair behaviors (which, unfortunately, are not unknown in the scientific setting), etc.

On the other hand, this problem is known to be structurally similar to that of ranking urls on the web. As is well known, this latter problem has been formalized in the {\em Google PageRank} \cite{BP98} (see also \cite{BP09}). Based on the simple idea that the importance of a web page depends on the number of web pages that link to it and on their relative importance, the {\em PageRank} relies on a solid mathematical basis which allows  the search engine Google \cite{google} to efficiently recover information across the web (see also \cite{SSC05,CS10} for a deeper mathematical analysis of the corresponding matrix problem, and \cite{BiDCRo089, BiDCRo10} for generalizations). Approaches based on this idea have been used for evaluating the impact of scientific journals (see, e.g., \cite{Be07,BVdSHC09}) and the impact of scientific articles (see, e.g., \cite{CXMR07,MR08,MGZ08,LW09}), also taking into account of collaborations \cite{DYFC09}. Sometimes, however, the above procedure
 s are not mathematically well refined.
In any case, such approaches use a {\em global information} which could be difficult to recover and manage efficiently (it is enough thinking to the computation of the {\em Google PageRank} to realize the possible complexity of the problem). Indeed, numerical algorithms need to be finely tuned, in order to gain efficiency (see, e.g., \cite{GG06,YYNg12}). This is the main reason why heuristics, like the number of citations (which are relatively easy to compute),  have become popular, even though, as pointed out above, sometimes they may provide misleading advices. Consequently, more efficient heuristics would be desirable for dealing with the problem.

With this premise, in Section~\label{due} we provide the model for constructing a mathematically grounded ranking of what we call the {\em PaperRank} problem\footnote{In analogy with the {\em PageRank} problem of the web.} which, under some mild assumptions, is proved to exist and to be unique. The results of this model are more reliable than those given by the model recently proposed in \cite{BiDCRo10}, and will be therefore assumed as ``reference solutions'' to validate a new heuristics, of local nature. The numerical examples provided in Section~\ref{tre} then clearly show that in many significant instances the new heuristics is much more reliable and fair than the usual one based on the bare number of citations.
In Section~\ref{quattro} the proposed model is  extended for ranking authors and journals in a straightforward way.
A few conclusions and remarks are then given in Section~\ref{last}.

\section{The {\em Random Reader} Model for the {\em PaperRank} Problem}\label{due}
The principle that we here describe is the analogous of that used in \cite{BP98} for deriving the famous {\em PageRank} of Google. For this reason, we name the problem {\em PaperRank} problem. We would like to remind that the mathematically based theory underlying the definition of the {\em Google PageRank} is the reason for its effectiveness in retrieving informations across the web. In the present setting, its basic principle may be then reformulated as follows:

\smallskip
\centerline{\em ``an important paper is cited by important papers.''}

\smallskip
\noindent That is, in analogy with the {\em random surfer} model proposed for the {\em PageRank} problem, we now have a virtual {\em random reader}, which starts reading a paper, then randomly passing to read a paper cited in it. If we repeat this process indefinitely, the {\em importance} of a given paper is the fraction of time that the random reader spends in reading it (assuming, obviously, that each paper is read in a constant time). This principle can be formally modeled by introducing the following {\em citation matrix}\,\footnote{Actually, by looking at the papers as the nodes of an oriented graph, such a matrix is nothing but the transpose of the adjacency matrix.}
\begin{equation}\label{L}
L = (\ell_{ij})\in\RR^{N\times N},\qquad
\ell_{ij} = \left\{ \begin{array}{cc} 1 &\mbox{~if paper $j$ cites paper $i$}\\ 0 &\mbox{otherwise}\end{array}\right., ~\forall i,j=1,\dots,N.\end{equation}
\begin{remark}
We observe that, by introducing the unit vector $$\bfe = (1,\dots,1)^T\in\RR^N,$$ then the vector containing the number of citations of each paper is given by
\begin{equation}\label{cit}
\bfc = \pmatrix{c} c_1\\ \vdots\\ c_N\endpmatrix \equiv L\,\bfe
\end{equation}
(i.e., $c_i$ is the number of citations of the $i$th paper). Such vector is currently used for computing several bibliometric indexes such as, e.g., the $H$-index.
\end{remark}
On the other hand, the vector $$\bff^T = (f_1,\dots,f_N) \equiv \bfe^TL $$ contains the number of bibliographic items in each paper. That is, $f_j$ is the number of references in paper $j$, $\forall j=1,\dots,N$. If we then define $v_i$ as the {\em importance} of the $i$th paper, then
$$v_i = \sum_{j=1}^N \ell_{ij}  v_j f_j^+, \qquad i=1,\dots,N,\qquad\quad f_j^+ =\left\{\begin{array}{cc} f_j^{-1} &\mbox{~if~} f_j>0\\ 0 &\mbox{~otherwise}\end{array}\right..$$
By introducing the vectors
 $$\bfv=(v_1,\dots,v_N)^T,  \qquad \bff^+ =(f_1^+\dots,f_N^+)^T,$$ and the matrices 
 \begin{equation}\label{FF}
 F = \diag(\bff), \qquad F^+=\diag(\bff^+),
 \end{equation} the previous set of equations can be cast in vector form as
\begin{equation}\label{S}
\bfv = LF^+\bfv \equiv S \bfv.
\end{equation} However, this ranking could not exist or might be not unique, depending whether $1\in\sigma(S)$, and/or if this eigenvalue is simple.

In order to cope with this problem, in \cite{BiDCRo10} the authors introduce, in a similar model, a {\em dummy paper}, say 0, which references all the other ones and is referenced by all of them.\footnote{We here consider only the problem of ranking the papers, whereas in \cite{BiDCRo10} a more general problem is modeled.} That is, matrix $L$ is replaced by the augmented matrix 
\begin{equation}\label{hL}
\hat{L} = \pmatrix{cc} 0 & \bfe^T\\ \bfe & L\endpmatrix\in\RR^{N+1\times N+1}.
\end{equation} Matrix $\hat{S}$ as in (\ref{S}) is then defined accordingly:
\begin{equation}\label{hS}
\hat{S} = \hat{L}\hat{F}^+\equiv \hat{L} \pmatrix{cc} N^{-1}\\ &(I+F)^{-1}\endpmatrix,
\end{equation}
with $F$ the diagonal matrix defined in (\ref{FF}) and $I$ the identity matrix of dimension $N$. Moreover, matrix (\ref{hS}) is clearly {\em irreducible}.

However, this last feature, makes the model not very faithful, in that it is quite well known that there exist groups of papers, whose citations do not overlap, so that matrix $L$ is indeed {\em reducible}. This is often the case, for example, for different fields of research within the same discipline or in different ones. For this reason, we here propose a different solution to this problem, in which we assume that, by default, {\em each paper references itself}, that is $\ell_{ii}=1$, for all $i=1,\dots,N$, so that (see (\ref{FF}))
$$f_j\ge1, \qquad j=1,\dots,N, \qquad \Longrightarrow \qquad F^+ = F^{-1}.$$
Consequently, $$\bfe^TS = \bfe^TL F^+ = \bff^T F^{-1} = \bfe^T, $$ so that $1\in\sigma(S)$ and, evidently, the possible reducibility of the original matrix (\ref{L}) is retained by the modified one.

Concerning the fact that ranking is unique, by following similar steps as those used for the {\em Google PageRank}, we may assume that, having reached a given paper, the {\em random reader} chooses with probability $p\in(0,1)$ a paper cited in it, or it {\em jumps} to read at random {\em any} paper, with probability $1-p$. In vector form this reads:
\begin{equation}\label{Sp}
\bfv = S(p) \bfv \equiv \left( pS +\frac{1-p}{N}\bfe \bfe^T\right)\bfv, \qquad p\in(0,1).
\end{equation}
Since $$S(p)>0, \qquad \|S(p)\|_1=1,\qquad \forall p\in(0,1), $$ from the Perron-Frobenius Theorem (see, e.g., \cite{Lanc}), one easily deduces that $1\in\sigma(S(p))$, which is a simple eigenvalue, separating in modulus all other eigenvalues of $S(p)$. In addition, the corresponding eigenvector $\bfv>0$. We then conclude that the {\em PaperRank} problem (\ref{Sp}) admits a solution, which is feasible (i.e., with positive entries) and unique.
Moreover, by choosing $p\approx1$, $S(p)\approx S(1)\equiv S$ and, therefore, the approximate model well matches the original one (see also Section~\ref{due.1} below).

Consequently, this ranking is {\em rigorously mathematically grounded}, though it requires an information of global nature, alike the case for computing the {\em Google PageRank}.
This means that it is relatively costly, since it requires to know all the data about every bibliographical item.

\begin{remark}\label{nop}
We observe that this last step (i.e., the introduction of the parameter $p$) is not required for the matrix $\hat{S}$ (see (\ref{hS})) of the model derived from (\ref{hL}), since one easily proves the following result.

\begin{theorem} 
Let $L\ne0$ and $\hat{S}$ defined according to (\ref{hS}). Then $\hat{S}^4>0$.
\end{theorem}
In other words, there always exists a path of exact length 4 between any two of the nodes $0,\dots,N$, provided that $L\ne0$.
\end{remark}

\subsection{Perturbation analysis}\label{due.1}
In this section we provide a simple analysis showing how the introduction of the parameter $p$ in (\ref{Sp}) affects the original vector. For this purpose, let us denote the eigenvector as $\bfv(p)$. That is,
$$S(p)\bfv(p) = \bfv(p), \qquad p\in(0,1).$$ Clearly, $\bfv^*\equiv \bfv(1)$ is the correct limit vector, which obviously exists, whereas $\bfv(0) = \frac{1}N\bfe$. Consequently, an estimate for $\bfv'(p)$ is given by $$\bfv' \approx \bfv(1)-\bfv(0) = \bfv^*-\frac{1}N\bfe.$$ One then obtains that, for $p\approx 1$:
\begin{equation}\label{apprv}
\bfv(p) \approx \bfv(1) + (p-1)\bfv' = p\bfv^*+\frac{1-p}N\bfe.
\end{equation}
From (\ref{apprv}) one then concludes that the introduction of the parameter $p$ results in an almost uniform (small) perturbation of the entries of the correct vector. As a matter of fact, the statistical properties of the two vectors are practically the same, for all the test problems reported in Section~\ref{tre}.

\subsection{A New Heuristics for the {\em PaperRank} Problem}\label{due.2}

As it has been shown in the previous section, the correct {\em PaperRank} is obtained by starting from the scaled matrix $LF^+$, in place of $L$. Similarly, instead of considering the bare number of citations, given by the vector (\ref{cit}), we propose to use {\em normalized} citations,  defined as the entries of the vector (see (\ref{S}))
\begin{equation}\label{citn}
\bfc_{norm} = LF^+\bfe \equiv S\bfe,
\end{equation}
which requires, as (\ref{cit}), only information of local nature. In other words, in place of counting the number of citation to a given paper, we propose to consider the {\em number of citations to that paper, each divided by the number of references in the corresponding paper containing the citation itself}. It is obvious that the index (\ref{citn}) has the same complexity as (\ref{cit}). Nevertheless, in  Section~\ref{tre} we show that the statistical properties of (\ref{citn}) are more fair than those of (\ref{cit}), in the sense that they better reproduce the correct ones provided by the reference model (\ref{Sp}).

It is evident, from the definition (\ref{citn}), that the vector $\bfc_{norm}$ is essentially the first iterate of the power method applied to $S$, by starting from a constant vector. Consequently,
\begin{itemize}
\item it requires only a {\em local} information, even though it would aim to approximate, in the limit, a global one (i.e., the {\em PaperRank});

\item no more than one iteration is possible, without requiring a global information.
\end{itemize}
Consequently, the heuristics (\ref{citn}) is the best we can do by using only local information. Nonetheless, as is shown in the numerical tests, it proves to be quite effective.

\begin{remark} It is worth mentioning that the use of the normalized citations (\ref{citn}) also copes correctly with the problem of self-citations. Indeed, for each new published paper, the normalized additional (self-)citations of an author cannot exceed 1. On the contrary, they are virtually unbounded for the vector (\ref{cit}) of bare citations.  

\end{remark}

\section{Numerical Tests}\label{tre}

We here provide a few numerical tests, each one modeling a significant situation, to compare the {\em PaperRank} obtained from (\ref{Sp}) by choosing $p=0.99$ (which we assume to be the reference one), with those given by the model (\ref{cit}), based on bare citations, and (\ref{citn}), based on the normalized citations. For each test we plot three histograms which rank the papers according to the vectors representing these indexes. For ease of a direct comparison, the obtained vectors are normalized so that their values range in the interval $[0,1]$.

Moreover, for each problem we also compare the {\em PaperRank} obtained from (\ref{Sp}) with that obtained from (\ref{hS}), that is from the model proposed in \cite{BiDCRo10}. In fact, we shall see that they may significantly differ, due to the fact that our model preserves the possible reducibility of the matrix (\ref{L}). On the other hand, it is clear that the corresponding vectors (\ref{cit}) and (\ref{citn}) derived 
by replacing the matrices $L$ and $S$ by the two matrices (\ref{S}) and (\ref{hS}) are essentially the same as the original ones (obviously, by neglecting the first entry of such vectors, i.e., that related to the {\em dummy paper}).

\subsection*{Examples 1 and 2}  
\begin{figure}[f]
\begin{center}
\includegraphics[width=8cm,height=6cm]{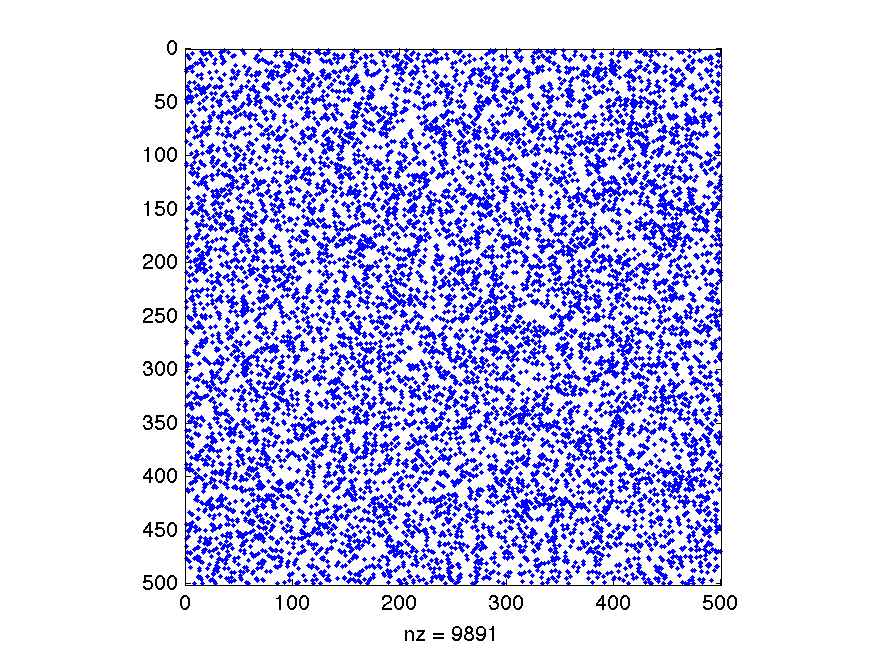}\\
\includegraphics[width=8cm,height=6cm]{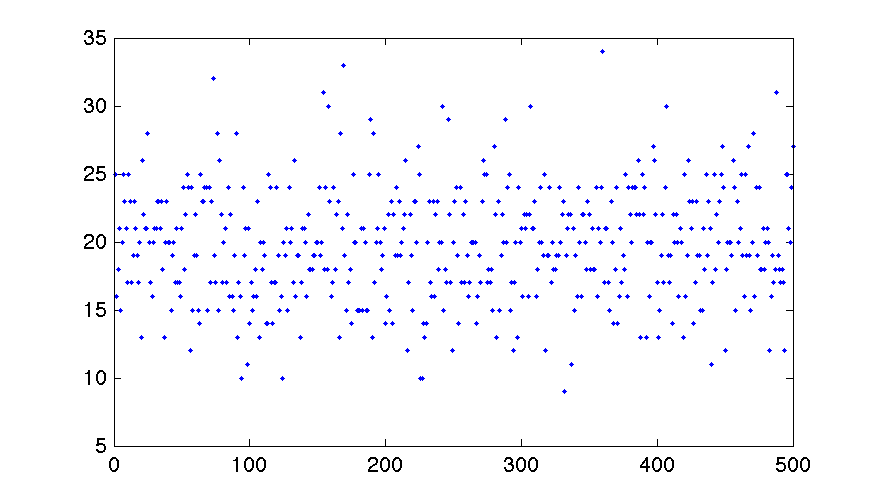}\\
\includegraphics[width=8cm,height=6cm]{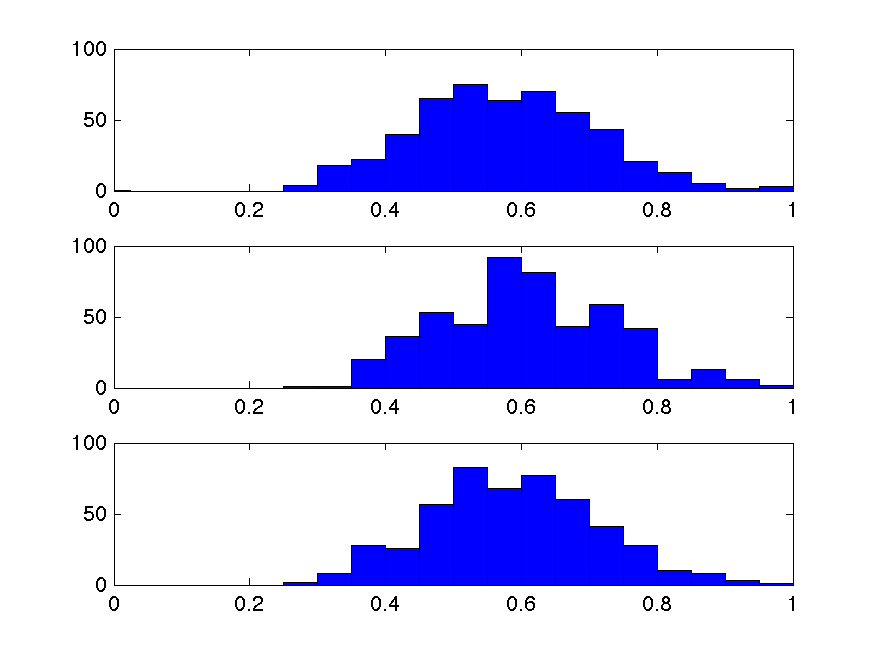}
\end{center}
\caption{citation matrix (\ref{L}), number of references (\ref{cit}), and rankings (\ref{Sp}), (\ref{cit}), and (\ref{citn}) for  Example~1, respectively.}\label{ex0}
\end{figure}

\begin{figure}[f]
\begin{center}
\includegraphics[width=8cm,height=6cm]{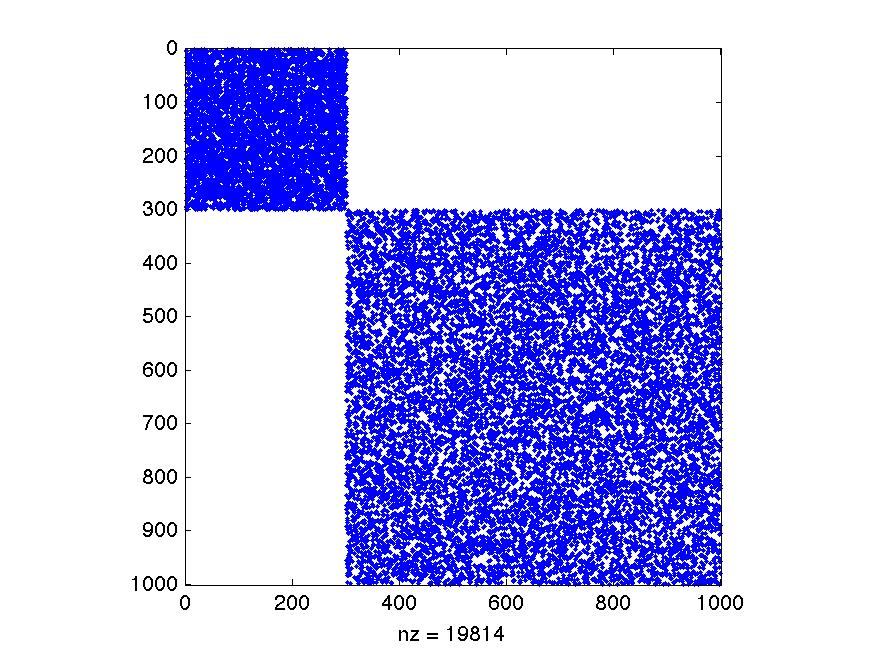}\\
\includegraphics[width=8cm,height=6cm]{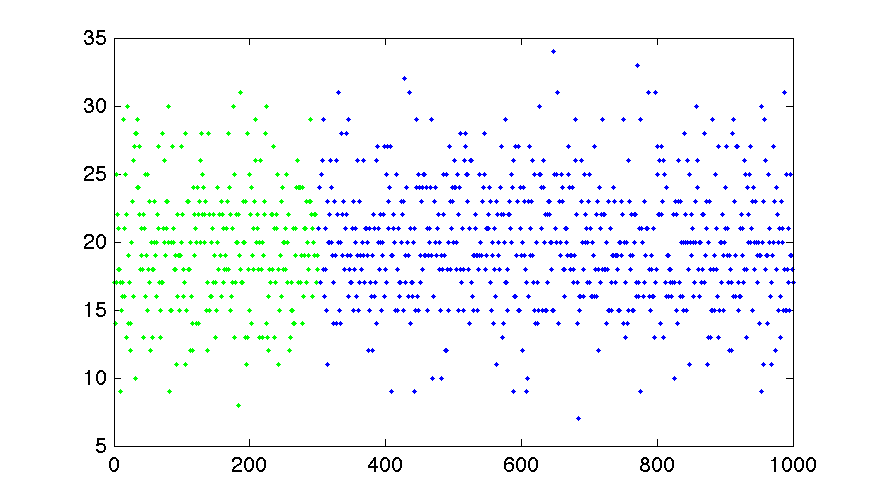}\\
\includegraphics[width=8cm,height=6cm]{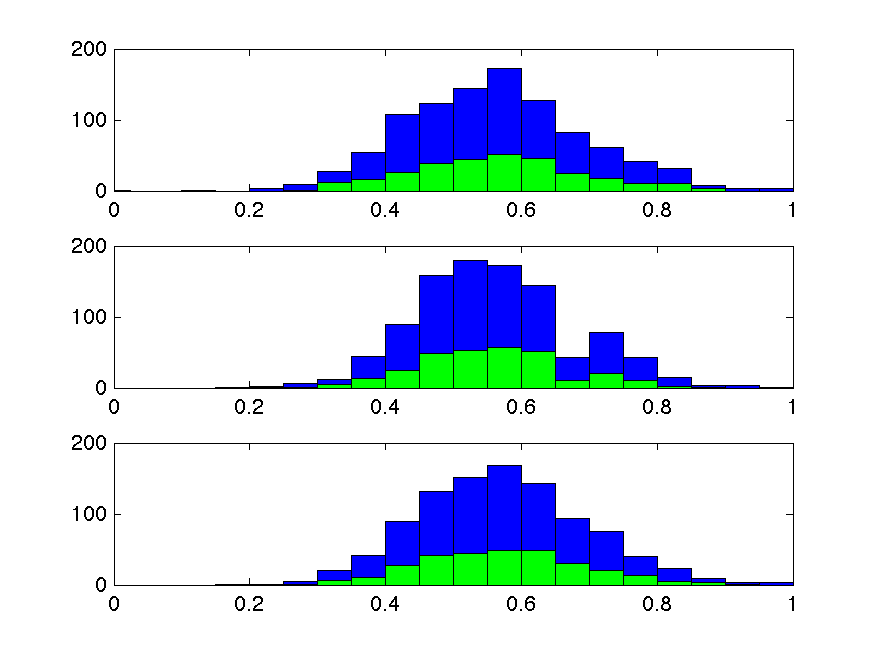}
\end{center}
\caption{citation matrix (\ref{L}), number of references (\ref{cit}), and rankings (\ref{Sp}), (\ref{cit}), and (\ref{citn}) for  Example~2, respectively.}\label{ex1}
\end{figure}

We suppose to have a single and homogeneous group of 500 articles (first example) or two distinct and homogeneous groups with 300 and 700 articles, respectively (second example). In both cases each paper has a mean of 20 randomly distributed references in its own group (see the first two plots in Figures~\ref{ex0} and \ref{ex1}). In both the examples, all the three rankings (\ref{Sp}), (\ref{cit}), and (\ref{citn}) have a similar distribution of the relevance of the papers, as is shown in the last three plots in Figures~\ref{ex0} and \ref{ex1}, even though (\ref{citn}) better fits the distribution of (\ref{Sp}). In Figure~\ref{ex1} the green bars concern the first group of papers, whereas the blue ones concern the second group.

For these problems, the {\em PaperRanks} obtained from (\ref{Sp}) and (\ref{hS}) turn out to be similar each other, so that we do not report the latter ones. Indeed, in the first example, the matrix is irreducible and in the second example reducibility is not an important feature, since the two blocks have similar properties (i.e., the same number of mean references in each paper).

\subsection*{Examples 3 and 4} 
\begin{figure}[t]
\begin{center}
\includegraphics[width=8cm,height=6cm]{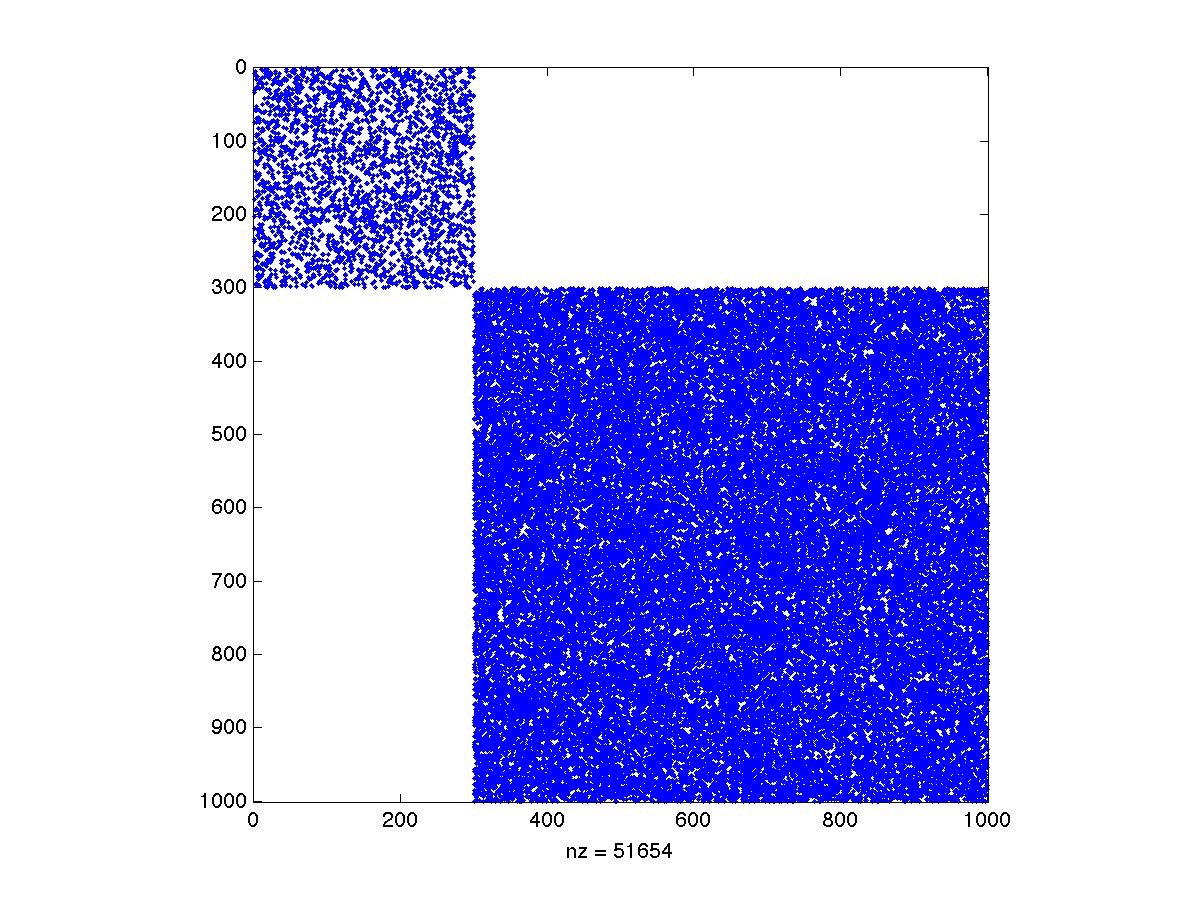}\\
\includegraphics[width=8cm,height=6cm]{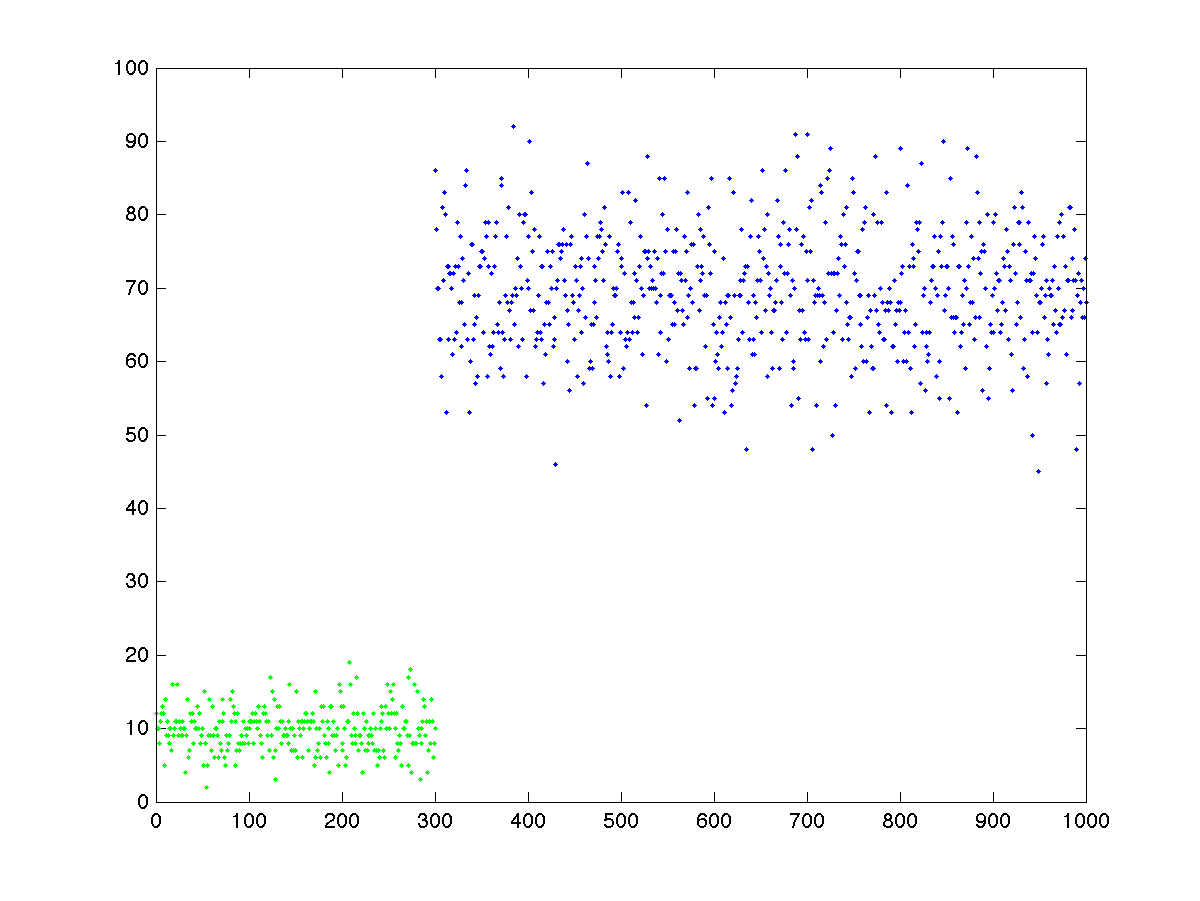}\\
\includegraphics[width=8cm,height=6cm]{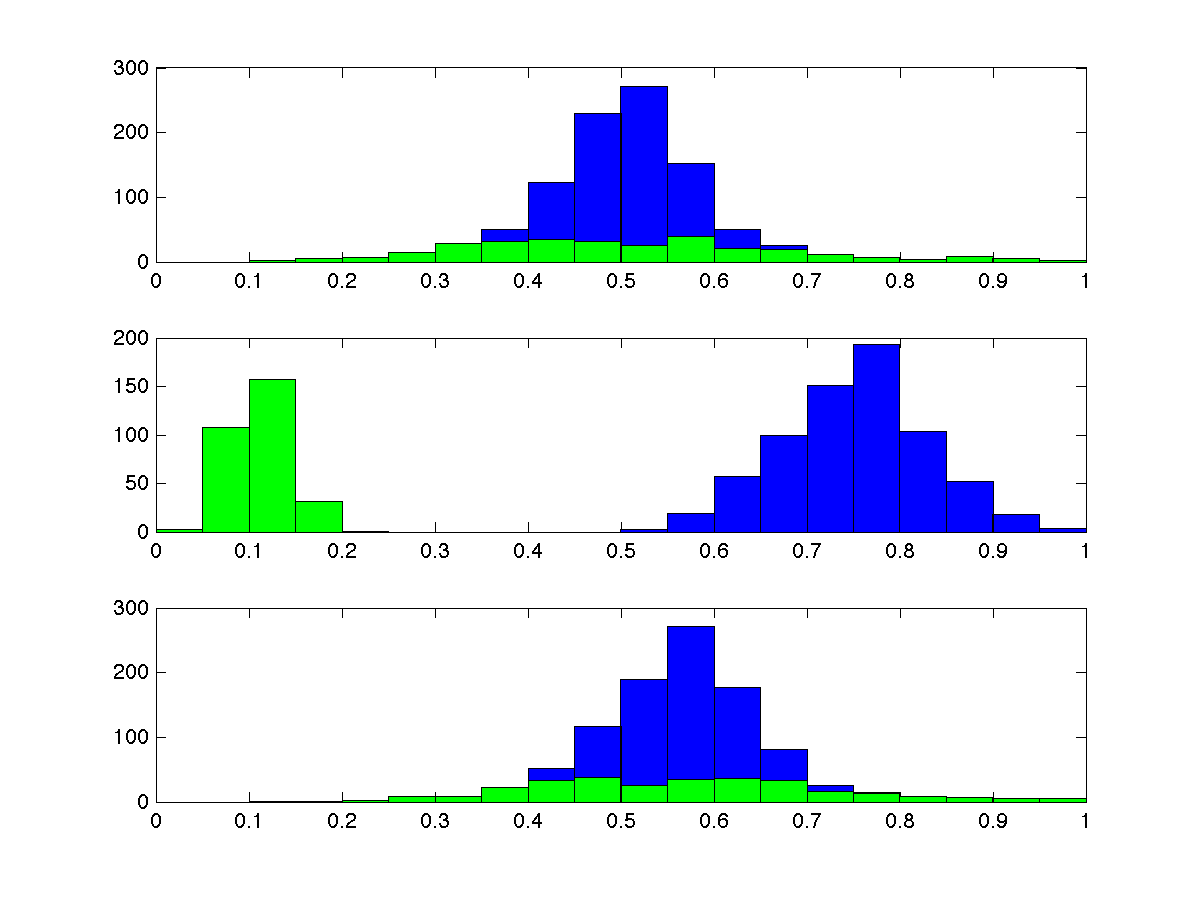}
\end{center}
\caption{citation matrix (\ref{L}), number of references (\ref{cit}), and rankings (\ref{Sp}), (\ref{cit}), and (\ref{citn}) for  Example~3, respectively.}\label{ex2}
\end{figure}

\begin{figure}[t]
\begin{center}
\includegraphics[width=8cm,height=6cm]{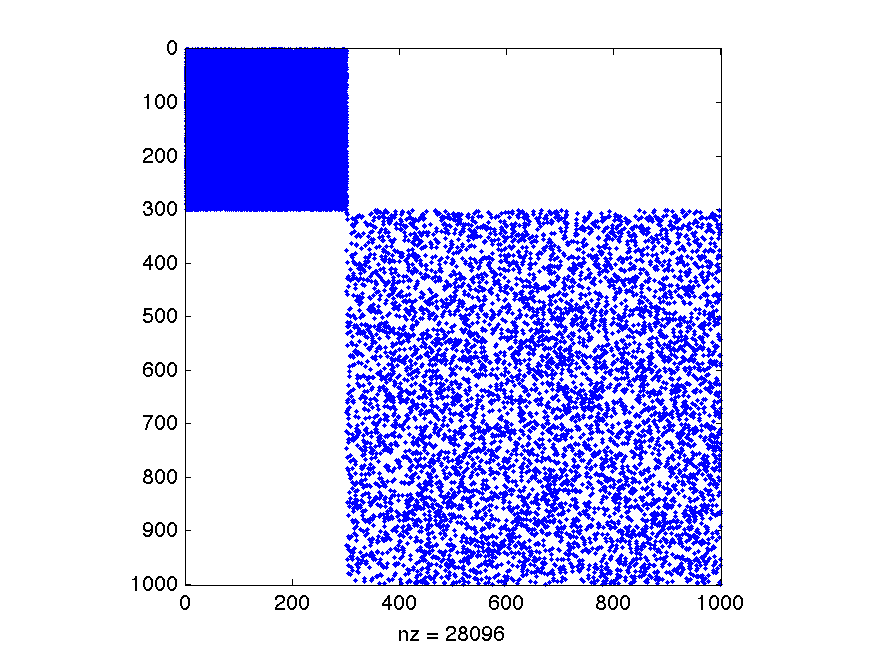}\\
\includegraphics[width=8cm,height=6cm]{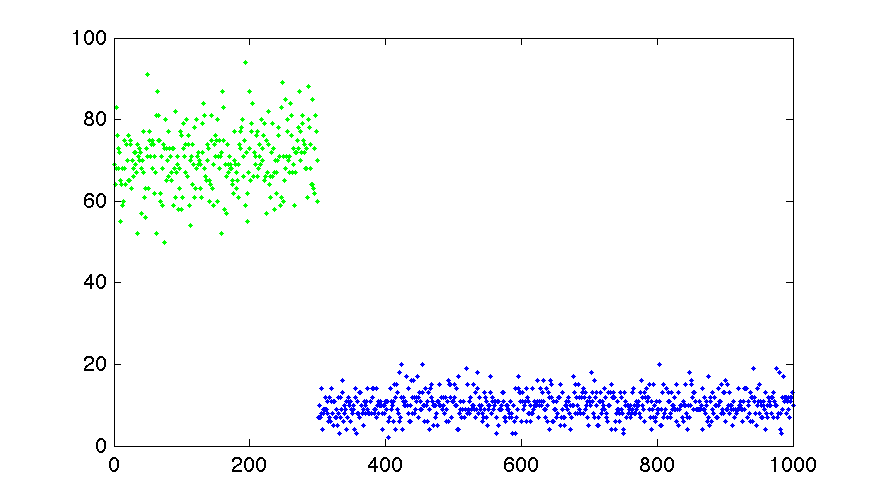}\\
\includegraphics[width=8cm,height=6cm]{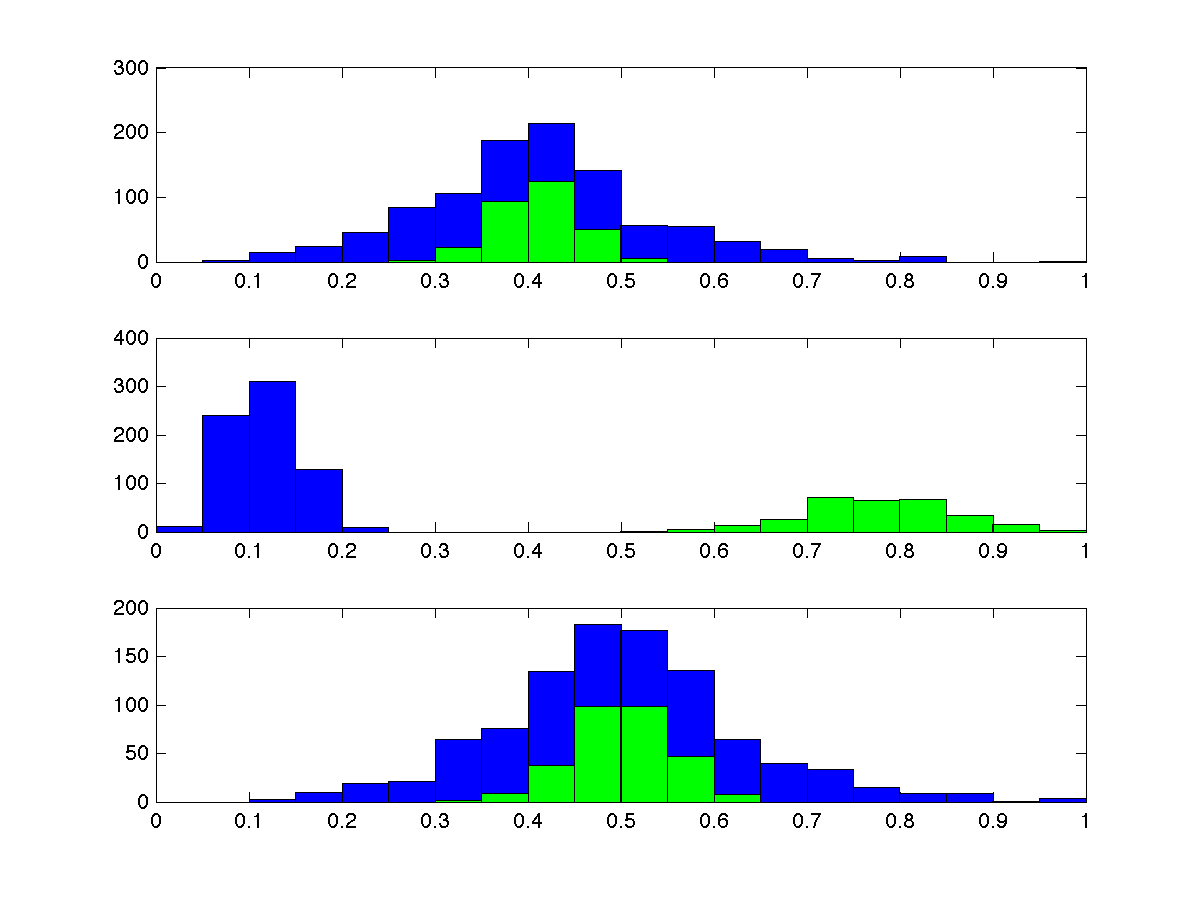}
\end{center}
\caption{citation matrix (\ref{L}), number of references (\ref{cit}), and rankings (\ref{Sp}), (\ref{cit}), and (\ref{citn}) for  Example~4, respectively.}\label{ex3}
\end{figure}

In these examples, we have two distinct and homogeneous groups of papers, with 300 and 700 items, respectively. Each paper only cites articles in its own group. In Example~3, each paper in the first group has a mean of 10 randomly distributed references, whereas each paper in the second group has a mean of 70 randomly distributed references (see the first two plots in Figure~\ref{ex2}). In Example~4, the situation is reversed since the papers in the smaller group have a mean of 70 randomly distributed citations whereas each paper in the second group has a mean of 10 randomly distributed references (see the first two plots in Figure~\ref{ex3}). In the last three plots of Figures~\ref{ex2} and \ref{ex3}, the green bars concern the first group of papers, whereas the blue ones concern the second group. The rankings (\ref{Sp}) and (\ref{citn}) always have a similar distribution of the relevance of the papers, whereas the ranking (\ref{cit}) exhibits 
 two peaks (one for the first group and one for the second group), which exchange in the two cases. It is clear that in these situations the different number of references in the papers of each group invalidate the ranking (\ref{cit}), whereas it doesn't affect the normalized ranking (\ref{citn}).

\begin{figure}[t]
\centerline{
\includegraphics[width=8cm,height=8cm]{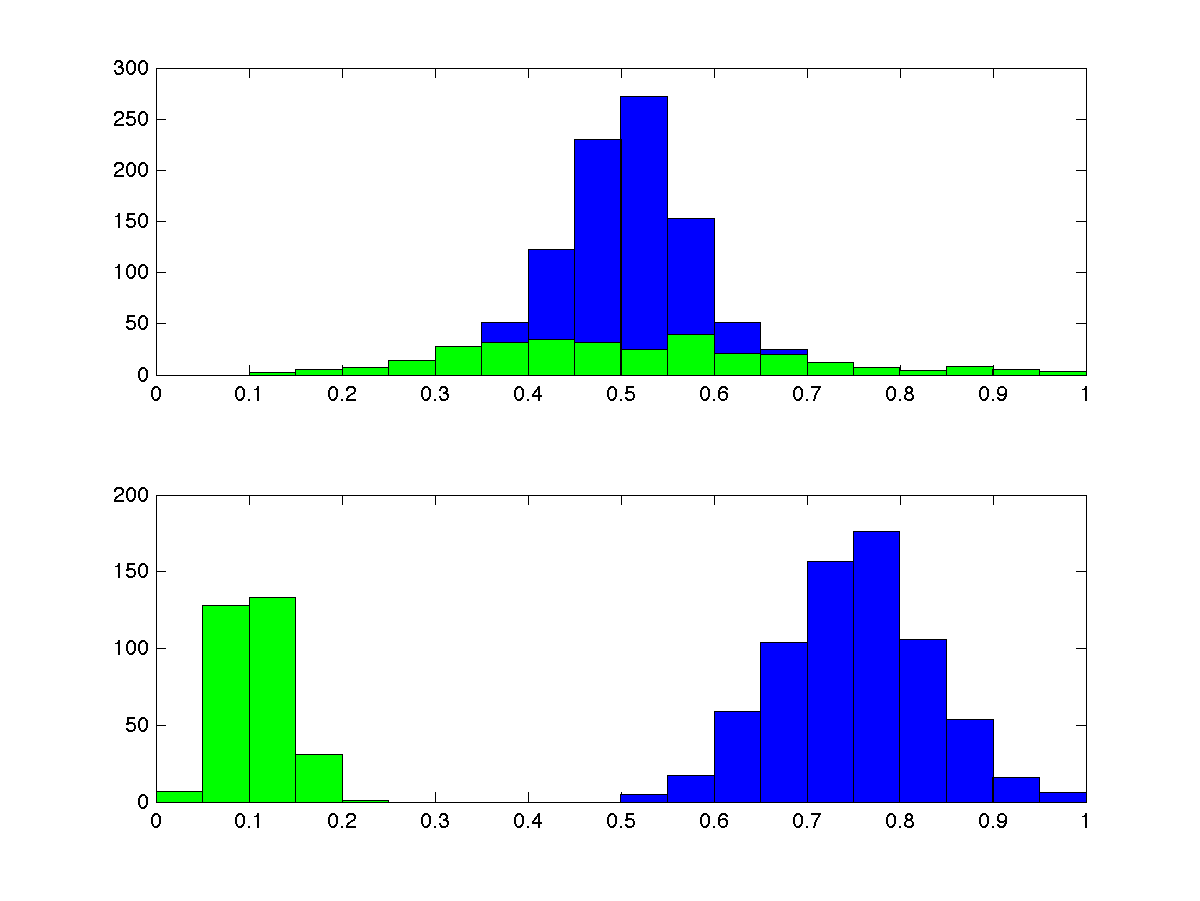}}
\caption{{\em PaperRanks} derived from (\ref{Sp}) (upper plot) and (\ref{hS})  (lower plot) for  Example~3.}\label{ppr1}

\bigskip\bigskip
\centerline{
\includegraphics[width=8cm,height=8cm]{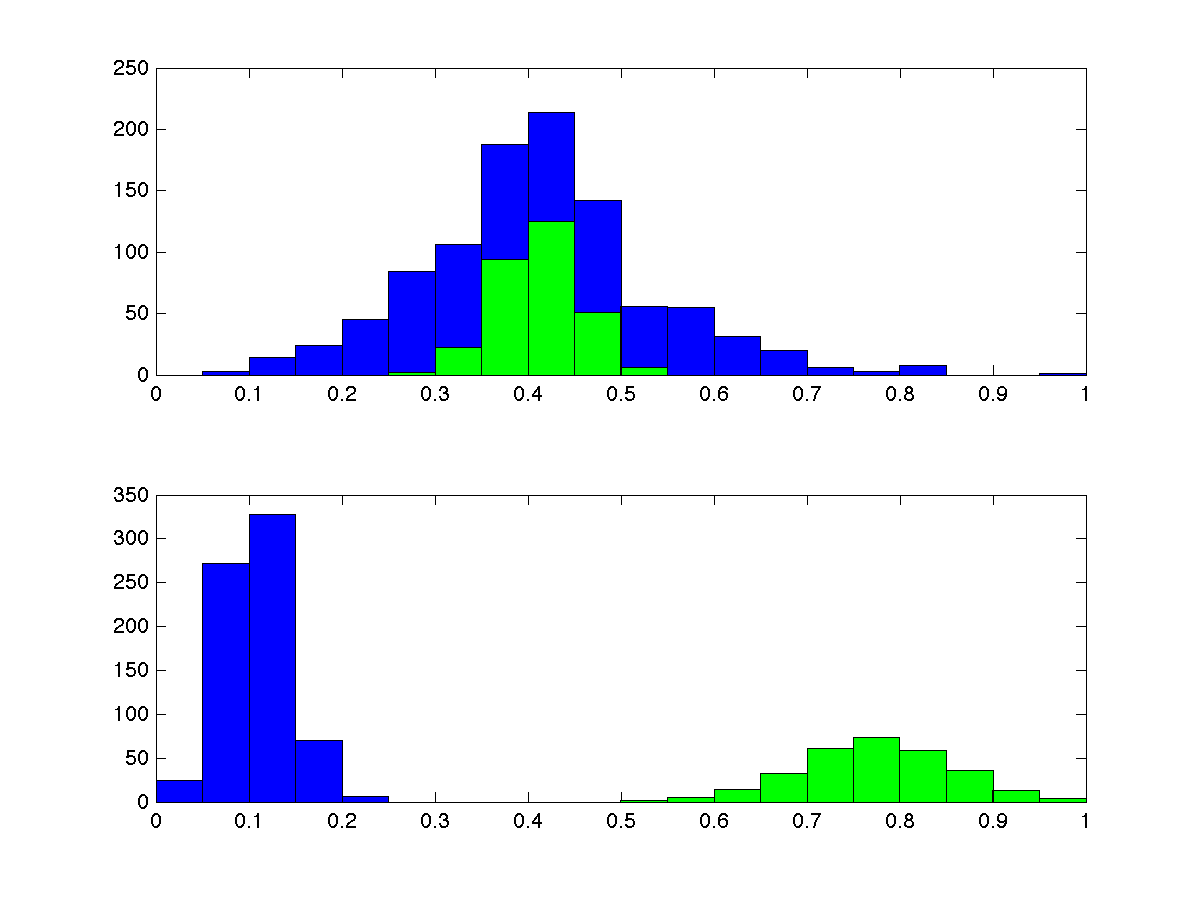}}
\caption{{\em PaperRanks} derived from (\ref{Sp}) (upper plot) and (\ref{hS})  (lower plot) for  Example~4.}\label{ppr2}
\end{figure}

As one may expect, for these problems, there is a significant difference between the {\em PaperRanks} obtained from our model (\ref{Sp}) and that derived from (\ref{hS}). Indeed, in both cases reducibility turns out to be an important feature of the corresponding citation matrices. This is shown in Figures~\ref{ppr1} and \ref{ppr2}, respectively, for the two examples. In each figure, the upper plot reproduces the third subplot in Figures~\ref{ex2} and \ref{ex3}, respectively, whereas the lower plot depicts the corresponding {\em PaperRank} derived from (\ref{hS}). It is evident that the latter one does not allow to properly compare the papers in the two groups.

\subsection*{Example 5} 
\begin{figure}[t]
\begin{center}
\includegraphics[width=8cm,height=6cm]{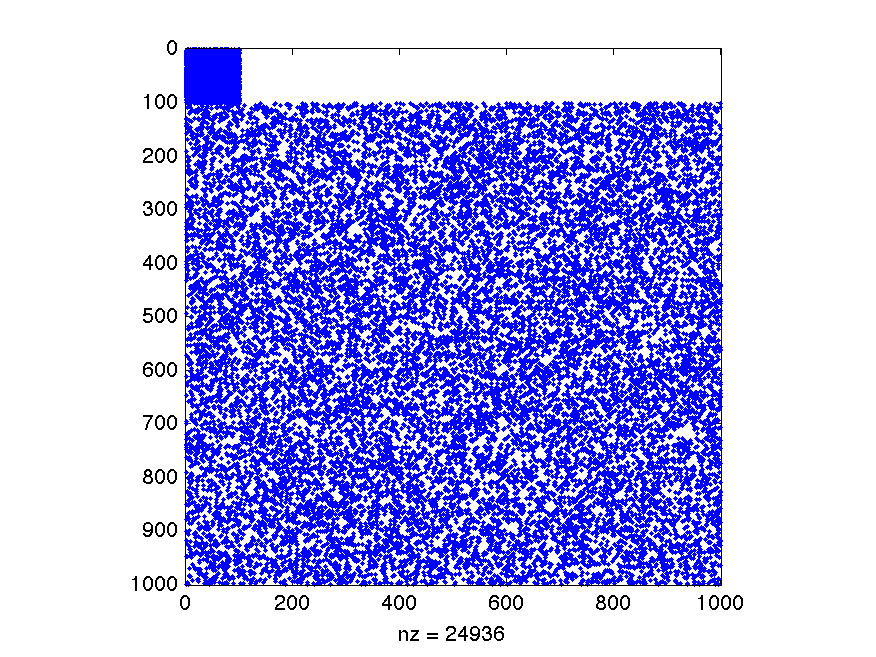}\\
\includegraphics[width=8cm,height=6cm]{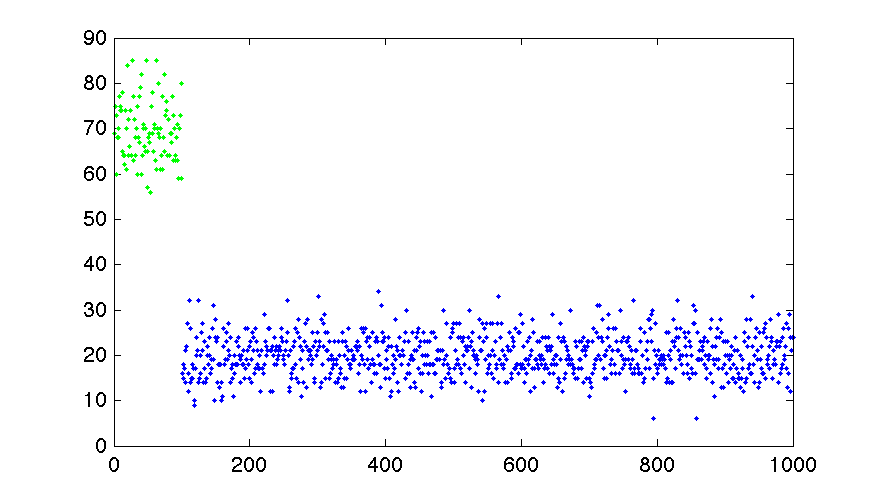}\\
\includegraphics[width=8cm,height=6cm]{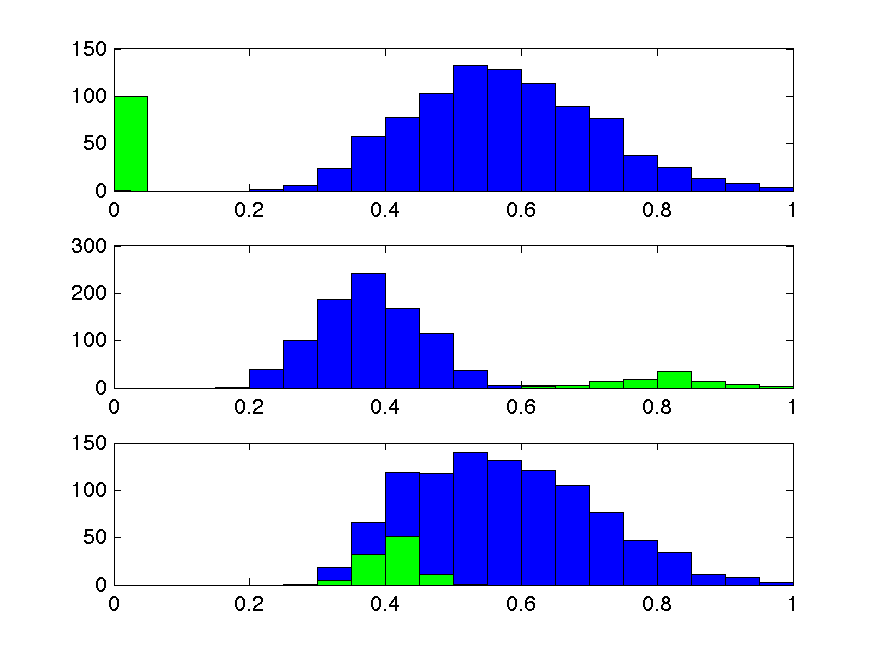}
\end{center}
\caption{citation matrix (\ref{L}), number of references (\ref{cit}), and rankings (\ref{Sp}), (\ref{cit}), and (\ref{citn}) for  Example~5, respectively.}\label{ex4}
\end{figure}

In this example, we have two groups of papers: a larger one, with 900 papers, which reference randomly a mean of 20 papers in the same group, and a smaller one of 100 papers, which reference papers in the same group, with a mean of 50 citations, and the papers in the larger one, with a mean of 20 citations (see the first two plots in Figure~\ref{ex4}). In this case, the heuristics (\ref{cit}), based on the bare number of citations, recognizes two groups of papers, the most important being the smaller one (the green bars in the fourth plot of Figure~\ref{ex4}). The correct distribution, however, is that depicted in the third plot of Figure~\ref{ex4}, given by (\ref{Sp}), where the green bars are the leftmost (i.e., the less important) ones. This behaviour is qualitatively better reproduced in the last plot of Figure~\ref{ex4}, concerning the heuristics (\ref{citn}).

For this problem, the {\em PaperRanks} obtained from (\ref{Sp}) and (\ref{hS}) turn out to be similar each other, since the citation matrix turns out to be irreducible. Consequently, we do not report the latter one.

\subsection*{Example 6} 
\begin{figure}[t]
\begin{center}
\includegraphics[width=8cm,height=6cm]{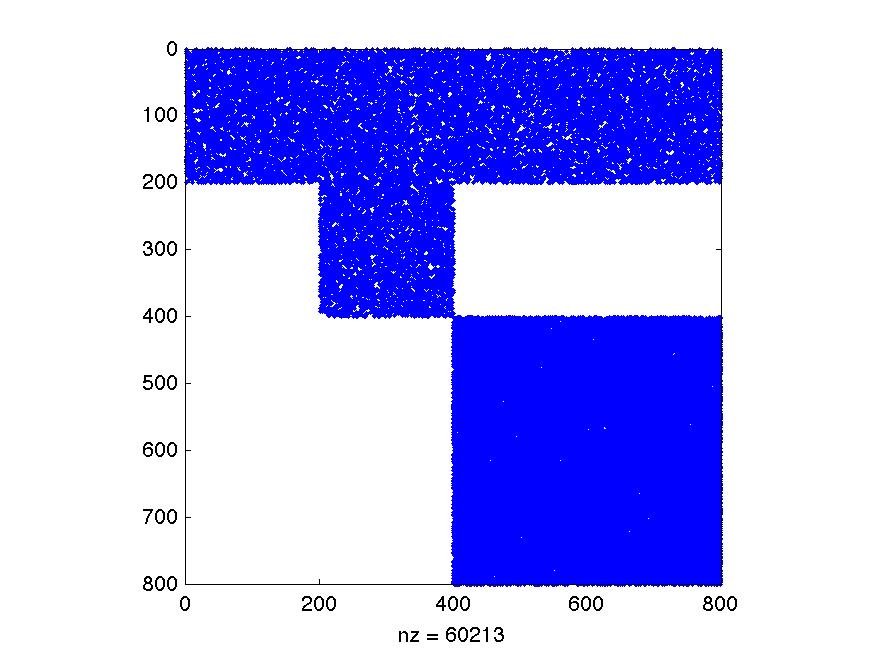}\\
\includegraphics[width=8cm,height=6cm]{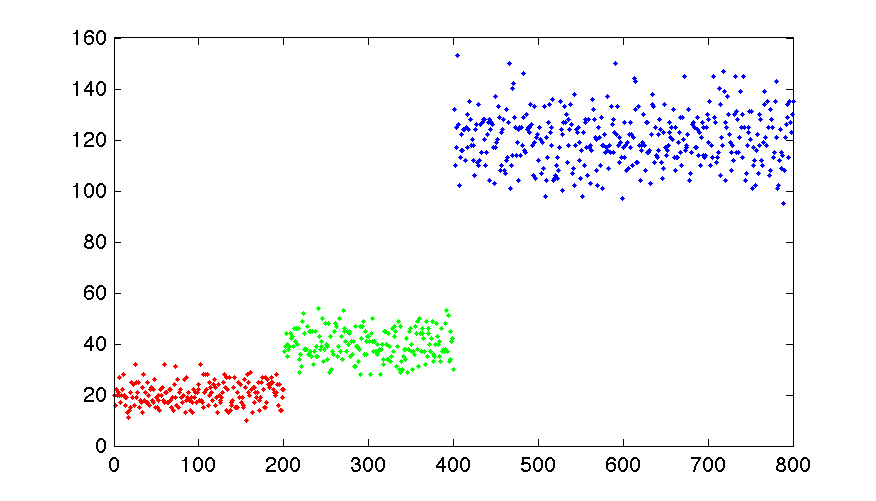}\\
\includegraphics[width=8cm,height=6cm]{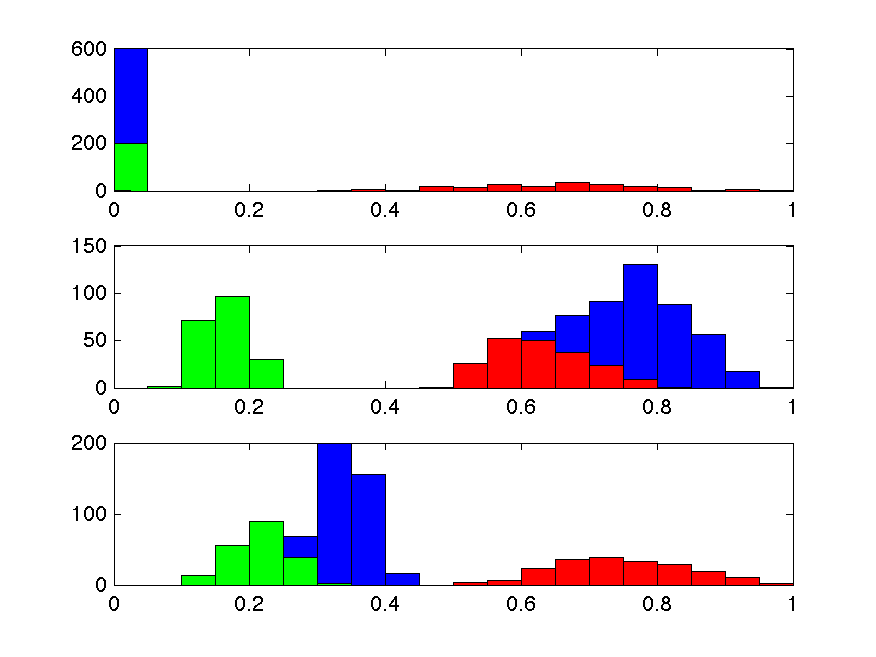}
\end{center}
\caption{citation matrix (\ref{L}), number of references (\ref{cit}), and rankings (\ref{Sp}), (\ref{cit}), and (\ref{citn}) for  Example~6, respectively.}\label{ex5}
\end{figure}

The last example concerns the case of three groups of papers:
\begin{itemize}
\item a group of 200 {\em leader} papers, which randomly reference a mean of 20 papers in the same group;

\item a group of 200 papers, which randomly reference a mean of 20 {\em leader} papers and 20 papers in its own group;

\item a group of 400 papers, which randomly reference a mean of 20 {\em leader} papers and 100 papers in its own group;
\end{itemize}
This situation is summarized by the first two plots in Figure~\ref{ex5}. It is evident that the correct ranking is that depicted in the third plot of Figure~\ref{ex5}, representing the vector in (\ref{Sp}), with the {\em leader} papers (in red) more important than those in the second group (in green) and those in the third group (in blue), these latter having the same importance. This situation is qualitatively well reproduced by the new heuristics (\ref{citn}), as is shown in the last plot of Figure~\ref{ex5}, where the {\em leader} papers (red) are again the most important, and the other  ones (green and blue) have a comparable importance, though the blue ones are slightly oversized. Vice versa, the usual ranking (\ref{cit}), based on the bare number of citations (which is shown in the fourth plot of Figure~\ref{ex5}), depicts a wrong scenario, in which the {\em leader} papers are replaced by those with the highest number of internal references (third group).

For this problem, the {\em PaperRanks} obtained from (\ref{Sp}) and (\ref{hS}) turn out to be similar each other, since the citation matrix turns out to be irreducible. Consequently,  we do not report the latter one.

\section{Ranking Authors and Journals}\label{quattro} 
Once papers are ranked (either exactly by using (\ref{Sp}) or approximately, by using the heuristics (\ref{citn})) it is possible to derive corresponding rankings for authors and journals, as we sketch below. Let us start with the ranking of authors.

Assume then that the rank $v_i$ of paper $p_i$ is known, for all $i=1,\dots,N$. Moreover,  for each paper $p_i$ let:
\begin{itemize}

\item $s_i$ be the set of all authors of paper $p_i$;

\item $n_i$ be the cardinality of $s_i$. I.e., the number of authors of paper $p_i$.
\end{itemize}
We then define the {\em rank of author} $a_j$ as:
$$r_j = \sum_{\{i\,:\,a_j\in s_i\}}  \frac{v_i}{n_i}, \qquad j=1,\dots,M,$$
where $M$ is the total number of distinct authors of all existing papers. Clearly, $r_j>0$, $j=1,\dots,M$. Moreover, it is an easy task to prove the following result.

\begin{theorem}\label{conserva1} One has: \begin{equation}\label{vieqrj}\sum_{i=1}^N v_i = \sum_{j=1}^M r_j.\end{equation}
\end{theorem}

\begin{rem}
In other words, (\ref{vieqrj}) means that the ``importance'' of all papers is fairly divided among all authors.  
\end{rem}

A similar approach can be used for ranking journals: the rank of journal $J_k$ is defined as
$$\ell_k = \sum_{\{i\,:\,p_i\in J_k\}} v_i, \qquad k=1,\dots,K,$$
where $K$ is the total number of journals. This means that the ``importance'' of a journal is given by the sum of the importance of the papers which have been published in it. Similarly as before, one has $\ell_k>0$, $k=1,\dots,K$. Moreover, the following result trivially holds.

\begin{theorem}\label{conserva2} One has: \begin{equation}\label{vieqlk}\sum_{i=1}^N v_i = \sum_{k=1}^K \ell_k.\end{equation}
\end{theorem}

\begin{rem}
Similarly as that stated for authors, (\ref{vieqlk}) means that the ``importance'' of all papers is exactly divided among the journals.  
\end{rem}

We observe that it could be easily defined a {\em $\nu$-years rank} of both authors and journals at a given time $t$, having set $d_i$ the date of publication of paper $p_i$, as 
$$r_j(t;\nu) = \sum_{\left\{i\,:\,a_j\in s_i\, \wedge\, d_i\ge t-\nu \right\}}  \frac{v_i}{n_i}, \qquad j=1,\dots,M,$$
and
$$\ell_k(t,\nu) = \sum_{\left\{i\,:\,p_i\in J_k\, \wedge\, d_i\ge t-\nu \right\}} v_i, \qquad k=1,\dots,K,$$
respectively.

We conclude this section, by observing that, for journals, it is appropriate to measure also the {\em average importance} of the published papers. That is, by setting $n_k$ the number of papers published in the journal $J_k$,
$$\bar{\ell_k} = \frac{\ell_k}{n_k}, \qquad k=1,\dots,K.$$ A corresponding {\em $\nu$-years average importance} is then defined as:
$$\bar{\ell}_k(t,\nu) = \frac{\ell_k(t,\nu)}{n_k(t,\nu)}, \qquad k=1,\dots,K,$$ having set $n_k(t,\nu)$ the number of papers published in the $\nu$ years before time $t$, in journal $J_k$.

\section{Conclusions}\label{last}

In this paper, we provide a mathematically correct definition of the {\em PaperRank} problem to assess scientific papers, which is able to properly compare also papers in disjoint groups, thus improving on a recent model proposed in \cite{BiDCRo10}.
On the basis of this new model, we provide a local heuristics, based on normalized citations, which appears to be quite effective (though much cheaper to compute), allowing to overcome some well known drawbacks of the ranking based on the bare number of citations. Moreover, the proposed {\em PaperRank} model allows to derive corresponding rankings for both authors and journals.

%\bibliographystyle{nonumber}

%%%%%%%%%%%%%%%%%
\end{document}